**Dynamic scaling theory of the forced translocation of a semi-flexible polymer through a nanopore**

Pui-Man Lam[*]
Physics Department, Southern University
Baton Rouge, Louisiana 70813

and

Yi Zhen[+]
Department of Natural Sciences, Southern University
6400 Press Drive
New Orleans, Louisiana 70126

**Abstract:**
We present a theoretical description of the dynamics of a semi-flexible polymer being pulled through a nanopore by an external force acting at the pore. Our theory is based on the tensile blob picture of Pincus in which the front of the tensile force propagates through the backbone of the polymer, as suggested by Sakaue and recently applied to study a completely flexible polymer with self-avoidance, by Dubbledam et al. For a semi-flexible polymer with a persistence length $P$, its statistics is self-avoiding for a very long chain. As the local force increases, the blob size starts to decrease. At the blob size $P^2/a$, where $a$ is the size of a monomer, the statistics becomes that of an ideal chain. As the blob size further decreases to below the persistence length $P$, the statistics is that of a rigid rod. We argue that semi-flexible polymer in translocation should include the three regions: a self-avoiding region, an ideal chain region and a rigid rod region, under uneven tension propagation, instead of a uniform scaling picture as in the case of a completely flexible polymer. In various regimes under the effect of weak, intermediate and strong driving forces we derive equations from which we can calculate the translocation time of the polymer. The translocation exponent is given by $\alpha = 1 + \nu$, where $\nu$ is an effective exponent for the end-to-end distance of the semi-flexible polymer, having a value between ½ and 3/5, depending on the total contour length of the polymer. Our results are of relevance for forced translocation of biological polymers such as DNA through a nanopore.

[*]puiman_lam@subr.edu;  [+]yzhen@suno.edu

## I. Introduction:

Driven polymer translocation through a nanopore has attracted much attention because of its involvement in crucially important biological processes and promising RNA and DNA sequencing technique[1-4]. Generally, most cells must transport macromolecules across membranes to function. It is believed that relatively thick molecules pass through nanometer-scale channels and the translocation of polynucleotides through proteic pores has been implicated in a variety of biological processes including phage infection, bacterial conjugation, uptake of oligonucleotides by certain organs, and transport across the nuclear envelope in plants.[5-8]. Experimental studies have been carried out by the insertion of a polymer into a pore of diameter comparable to the size of the chemical repeat units that make up the polymer. For example, Kasianowicz, et al [8] have detected single strands of RNA passing through a 1.5 nm pore formed by a membrane-bound protein. The experimental results show that the ability of the polymer to enter the nanopore depends linearly on polymer concentration, and translocation time is highly sensitive to the polynucleotide sequence of ssDNA and RNA, and secondary structure of RNA [6-11].

Considerable theoretical attempts have been made to interpret experimental observations of dynamics of driven polymer translocation, using the scaling technique which achieved great success with models in which the polymer is regarded as a completely flexible, uniform "string" whose conformational entropy dominates the system's behavior, fractional Brownian motion which governs the dynamics of a single polymer moving through a nanopore in a membrane and Fokker-Planck equation of motion which contains time-dependent drift and diffusion terms [12-15].

Recently, Dubbeldam et al. [16] suggested a consistent generalization of the tensile force propagation model based on the tensile (Pincus) bloc picture and the notion of a propagating front of tensile force along the chain backbone [17-21]. Their model, however, is only applicable to completely flexible polymers and therefore would not apply to semi-flexible polymers, such as double-stranded DNA which have large persistence lengths of the order of 50 nm. Inspired largely by this work, in the present paper we develop the theoretical treatment for the translocation of a semi-flexible polymer through a nanopore. We argue that semi-flexible polymer in translocation should include a self-avoiding region, random walk region and a rigid rod region under uneven tension propagation instead of a uniform scaling picture. We propose that for a semi-flexible polymer under a sufficiently strong external force, two extra propagation fronts $X_Q$, $X_p$ can be defined where the tensile blob size is equal to $P^2/a$ and $P$ respectively, where $P$ is the persistence length and $a$ is the monomer size. At weak local force, the blob size is larger than $P^2/a$ and the monomers inside such a blob obey self-avoiding statistics. Closer to the pore, the local force increases and the blob size decreases to below $P^2/a$. The monomers inside such a blob obey ideal chain statistics [22,23]. Even closer to the pore, the local force can be strong enough so that the blob size is below the persistence $P$. Inside such a blob the monomers behave like a rigid rod. Our approach follows ref. [16] in formulating the dynamics of various regimes ("trumpet", "stem-trumpet", and "stem" regimes). We differ, however, in introducing the roles that the persistence length and the extra propagation fronts $X_Q$ and $X_p$ play in the translocation process. The translocation exponent is given by $\alpha = 1+\nu$, where $\nu$ is an effective exponent for the end-to-end distance of the semi-flexible polymer, having a value between ½ and 3/5, depending on the total contour length of the polymer.

    The organization of the paper is as follows: In Sec. II we describe the model and

investigate the dynamics of the polymer in various regimes under the effect of weak, intermediate and strong driving forces. For each force regime, we derive equations from which we can calculate analytically the translocation time of the polymer. Sec III is the conclusion.

**II. The Model**

We assume that the pulling is applied only to the monomer, which is inside the pore. As the pulling force is switched on, tension propagates along the backbone of the polymer and alters its conformation progressively. In Figure 1 we show the conformation of the polymer for three different situations of the external pulling force: (i) Weak force: $Ta/P^2 < f \leq T/P$, (ii) Intermediate force: $T/P < f < T/a$ and (iii) strong force: $f > T/a$. Here $P$ is the persistence length of the polymer, $T$ is the temperature, $a$ is the microscopic length scale of a monomer ( for DNA this is the bp-bp separation of 0.34 nm ), $N_{tot}$ is the total number of monomers in the polymer, $f$ is the applied force at the pore. Since the derivation of the equations are similar in the above three force regimes, we will give detailed derivations only for the intermediate force regime (ii) where blobs of all sizes (from greater than $P^2/a$ to less than $P^2/a$ and to less than $P$ ) can occur and summarize the results for the weak force and strong force regimes.

        1. Intermediate force: $T/P < f \leq T/a$

We will consider the case when $f > T/P$ where $P$ is the persistence length. For $f > T/P$, a blob of size less than or equal to $P$ can be formed, in addition to the blob of size $P^2/a$. Inside blobs of size smaller than $P^2/a$ but greater than $P$, the monomers obey ideal chain statistics while inside the blob of size smaller than $P$, the monomers behave like a rigid rod. This is the stem-trumpet regime illustrated in Figure 1b. The polymer is pulled by fixed external force $f$ at the position of the nanopore located at $x = 0$. The blob size $\mathbf{x}(x,t)$ at position x and time t is given by

$$\mathbf{x}(x,t) = \frac{T}{f_{loc}(x,t)} \qquad (1)$$

where $f_{loc}(x,t)$ is the local force at that position and time. We will follow the initial analysis of references [16,20] in order to generalize it to study the case of a semi-flexible polymer with a finite persistence length.

Let $-X(t)$ denote the position from the nanopore to which the tension has propagated at time t, i.e. there is no tension on the chain for position $x < -X(t)$. Then the local force $f_{loc}(x,t)$ is given by the Stokes friction force acting on the segments located between $x$ and $-X(t)$:

$$f_{loc}(x,t) = 6p\mathbf{z}_0 \int_{-X(t)}^{x} v(x',t) \left[\frac{\mathbf{x}(x',t)}{a}\right]^{z-2} \frac{dx'}{\mathbf{x}(x',t)} \qquad (2)$$

In this equation, $dx'/\mathbf{x}(x',t)$ counts the number of blobs in the interval $x', x'+dx'$ while $6p\mathbf{z}_0[\mathbf{x}(x',t)/a]^{z-2}$ is the local Stokes friction force, with $\mathbf{z}_0$ the friction coefficient [14]. z is the dynamic exponent and is equal to 3 or $2+1/\mathbf{n}$ for Rouse or Zimm dynamics respectively, where $\mathbf{n}$ is the end-to-end distance exponent. For convenience, in this work we will use the Flory value $\mathbf{n} = 3/5$. Also we will limit ourselves to Rouse dynamics only with z=3. The moving domain can be treated as a uniform block within which all monomers are moving with the same time-dependent velocity. Then, using Eqn. (2), Eqn.

(1) becomes,

$$\frac{1}{\mathbf{x}(x,t)} = \frac{6\mathbf{pz}_0 v(t)}{T} \int_{-X(t)}^{x} \left[\frac{\mathbf{x}(x't)}{a}\right]^{z-2} \frac{dx'}{\mathbf{x}(x't)} \qquad (3)$$

Differentiating this with respect to x, we obtain

$$\frac{d\mathbf{x}}{dx} = -\frac{6\mathbf{pz}_0 v(t)}{Ta^{z-2}} \mathbf{x}^{z-1} \quad . \qquad (4)$$

The boundary condition at $x = 0$ is given by

$$\mathbf{x}(x=0,t) = \frac{T}{f} \qquad (5)$$

as follows from Eqn. (1) because at x=0, the local force is the just external applied force $f$. The solution of (4) satisfying the boundary condition (5) is

$$\mathbf{x}(x,t) = \frac{a}{\left\{\frac{6\mathbf{pz}_0 v(t)}{T} x + \left[\frac{af}{T}\right]^{z-2}\right\}^{1/(z-2)}} \qquad (6)$$

The second boundary condition at $x = -X(t)$ is obtained from Eqn. (1) with $f_{loc}(x=-X(t),t) = 0$, or $\mathbf{x}(-X(t),t) = \infty$. This gives

$$-\frac{6\mathbf{pz}_0 v(t) X(t)}{T} + \left[\frac{af}{T}\right]^{z-2} = 0 \qquad (7)$$

Defining dimensionless quantities $\tilde{v}(t) = 6\mathbf{pz}_0 av(t)/T$, $\tilde{f} = af/T$, $\tilde{X}(t) = X(t)/a$, $\tilde{\mathbf{x}} = \mathbf{x}/a$ and $\tilde{x} = x/a$, Eqns. (7) and (6) can be rewritten as

$$\tilde{v}(t) = \frac{\tilde{f}^{z-2}}{\tilde{X}(t)} \qquad (8)$$

$$\tilde{\boldsymbol{x}}(\tilde{x},t) = \frac{1}{\left\{\tilde{v}(t)\left[\tilde{x}+\tilde{X}(t)\right]\right\}^{1/(z-2)}} \qquad (9)$$

Now for a semi-flexible polymer with a persistence length $P$, we can introduce the quantity $X_Q(t) < X(t)$, such that the blob size $\boldsymbol{x}(-X_Q(t),t)$ becomes equal to $P^2/a$:

$$\boldsymbol{x}(-X_Q(t),t) = P^2/a \qquad (10)$$

In terms of the dimensionless quantities $\tilde{P} = P/a$ and $\tilde{X}_Q = X_Q/a$, this becomes

$$\tilde{\boldsymbol{x}}(-\tilde{X}_Q(t),t) = \tilde{P}^2 \qquad (11)$$

Substituting this into Eqn. (9), we obtain an expression for $\tilde{X}_Q$:

$$\tilde{X}_Q = \tilde{X} - \frac{1}{\tilde{P}^{2(z-2)}\tilde{v}} \qquad (12)$$

In addition to the quantity $X_Q$ we have to introduce another quantity $X_P(t) < X_Q < X(t)$, such that the blob size $\boldsymbol{x}(-X_P(t),t)$ becomes equal to the persistence length:

$$\boldsymbol{x}(-X_P(t),t) = P \qquad (13)$$

In terms of the dimensionless quantities $\tilde{P} = P/a$ and $\tilde{X}_P = X_P/a$, this becomes

$$\tilde{\boldsymbol{x}}(-\tilde{X}_P(t),t) = \tilde{P} \qquad (14)$$

Substituting this into Eqn.(9), and using $z = 3$ we obtain an expression for $\tilde{X}_P$:

$$\tilde{X}_P = \tilde{X} - \frac{1}{\tilde{P}\tilde{v}} \qquad (15)$$

Let $M(t)$ denote the number of translocated monomers and $N(t)$ denote the total number of monomers subject to tension during the time interval [0,t]. Then the global material balance of monomers would require

$$\int_{-X(t)}^{-X_Q(t)} \left[\frac{\mathbf{x}(x,t)}{a}\right]^{1/n} \frac{dx}{\mathbf{x}(x,t)} + \int_{-X_Q(t)}^{-X_P(t)} \left[\frac{\mathbf{x}(x,t)}{a}\right]^2 \frac{dx}{\mathbf{x}(x,t)} + \int_{-X_P(t)}^{0} \left[\frac{\mathbf{x}(x,t)}{a}\right] \frac{dx}{\mathbf{x}(x,t)} + M(t) = N(t)$$

(16)

We have assumed that for $-X(t) \leq x \leq -X_Q(t)$, the blob size is larger than $P^2/a$ and the monomers inside the blob obey self-avoiding walk statistics, with monomer number given by $[\mathbf{x}/a]^{1/n}$. For $-X_Q(t) \leq x \leq -X_P(t)$, the blob size is smaller than $P^2/a$ and the monomers inside the blob obey ideal chain statistic, with monomer number given by $[\mathbf{x}/a]^2$. For $-X_P(t) \leq x \leq 0$, the blob size is smaller than $P$ and the monomers inside the blob behave like a rigid rod, with monomer number given by $[\mathbf{x}/a]$. In terms of the dimensionless quantities, Eqn. (16) can be written as

$$\int_{-\tilde{X}(t)}^{-\tilde{X}_Q(t)} [\tilde{\mathbf{x}}(x,t)]^{\frac{1}{n}-1} d\tilde{x} + \int_{-\tilde{X}_Q(t)}^{-\tilde{X}_P(t)} \tilde{\mathbf{x}}(x,t) d\tilde{x} + \tilde{X}_P(t) + M(t) = N(t) \qquad (17)$$

Substituting Eqn. (9) into this, the first term becomes

$$\int_{-\tilde{X}(t)}^{-\tilde{X}_Q(t)} [\tilde{\mathbf{x}}(x,t)]^{\frac{1}{n}-1} d\tilde{x} = \tilde{v}^{\left(-\frac{1}{n}+1\right)/(z-2)} \int_{-\tilde{X}(t)}^{-\tilde{X}_Q(t)} (\tilde{x} + \tilde{X})^{(-\frac{1}{n}+1)/(z-2)} d\tilde{x} \qquad (18)$$

Using $z = 3$, and Eqns.(8), (9) and (12) become,

$$\tilde{v}(t) = \frac{\tilde{f}}{\tilde{X}(t)} \qquad (19)$$

$$\tilde{x}(\tilde{x},t) = \frac{1}{\tilde{v}(t)[\tilde{x} + \tilde{X}(t)]} \qquad (20)$$

$$\tilde{X}_Q = \tilde{X} - \frac{1}{\tilde{P}^2 \tilde{v}} \qquad (21)$$

Using Eqns. (19)-(21) Eqn. (18) becomes,

$$\int_{-\tilde{X}(t)}^{-\tilde{X}_Q(t)} [\tilde{x}(x,t)]^{\frac{1}{n}-1} d\tilde{x} = \tilde{v}^{-\frac{1}{n}+1} \int_{-\tilde{X}(t)}^{-\tilde{X}_Q(t)} (\tilde{x} + \tilde{X})^{-\frac{1}{n}+1} d\tilde{x} = \frac{1}{(2-1/n)\tilde{v}\tilde{P}^{2(2-1/n)}} = \frac{\tilde{X}}{(2-1/n)\tilde{f}\tilde{P}^{2(2-1/n)}} \qquad (22)$$

Using Eqns. (19)-(21) and (15) the second term in Eqn. (17) becomes

$$\int_{-\tilde{X}_Q(t)}^{-\tilde{X}_P(t)} \tilde{x}(x,t) d\tilde{x} = \int_{-\tilde{X}_Q(t)}^{-\tilde{X}_P(t)} \frac{d\tilde{x}}{\tilde{v}[\tilde{x}+\tilde{X}(t)]} = \frac{\tilde{X}}{\tilde{f}} \ln(\tilde{P}) \qquad (23)$$

At time $t = 0$, the external force is zero and all $N(t)$ monomers have occupied the region of size $X(t)$. Therefore $N(t)$ must satisfy the condition

$$N(t) = [\tilde{X}(t)]^{1/m} \qquad (24)$$

where $m$ is the effective exponent for the end-to-end distance of the semi-flexible polymer, having a value between ½ and 3/5, depending on the total contour length of the polymer. If the total contour length of the semi-flexible polymer is less than or equal to $P^2/a$, $m$ will be close to ½. If it is larger than $P^2/a$, $m$ will be close to 3/5. For DNA, $P^2/a \approx (150)^2 a \approx 22500a$.

Substituting Eqns. (22) and (23) into Eqn. (17) and using expression (24) for $N(t)$ we

obtain

$$\frac{\tilde{X}}{(2-1/n)\tilde{f}\tilde{P}^{2(2-1/n)}} + \frac{\tilde{X}}{\tilde{f}}\ln(\tilde{P}) + \tilde{X}_P + M(t) = \tilde{X}^{1/m} \qquad (25)$$

The flux of monomers at the pore is $j_0(t) = r(x=0,t)v(t)$, where $r(x,t) = [x(x,t)/a]^{1/n}/x(x,t)$ is the linear density of monomers. But at the pore ($x=0$), where the force is the strongest, we have assumed that the monomers inside this blob are in the rigid rod state. Therefore $r(0,t) = [x(0,t)/a]/x(0,t) = 1/a$. The flux $j_0(t)$ should be equal to $dM(t)/dt$, i.e.

$$\frac{dM(t)}{dt} = \frac{1}{a}v(t) = \frac{\tilde{v}T}{6pz_0 a}, \qquad (26)$$

where $\tilde{v}(t) = 6pz_0 av(t)/T$ was defined before. Using Eqn.(6) with $z=3$, Eqn. (19), and introducing the dimensionless time $\tilde{t} = t/t_0$, with $t_0 = a^2 z_0/T$, this becomes

$$\frac{dM(\tilde{t})}{d\tilde{t}} = \frac{\tilde{f}}{6p\tilde{X}(\tilde{t})} \qquad (27)$$

Differentiating Eqn.(25) with respect to $\tilde{t}$ and using Eqn. (27), we obtain

$$\left[\frac{1}{(2-1/n)\tilde{f}\tilde{P}^{2(2-1/n)}} + \frac{1}{\tilde{f}}\ln(\tilde{P}) + \left(1 - \frac{1}{\tilde{P}\tilde{f}}\right) - \frac{\tilde{X}^{1/m-1}}{m}\right]\frac{d\tilde{X}}{d\tilde{t}} = -\frac{\tilde{f}}{6p\tilde{X}} \qquad (28)$$

With the natural initial condition $\tilde{X}(0) = 1$, Eqn. (28) can be solved to yield

$$\tilde{t} = \tilde{t}_0 - \frac{3p}{\tilde{f}^2}\left[\frac{1}{(2-1/n)\tilde{P}^{2(2-1/n)}} + \ln(\tilde{P}) + \tilde{f}\left(1-\frac{1}{\tilde{P}\tilde{f}}\right)\right]\tilde{X}^2 + \frac{6p\tilde{X}^{1+1/m}}{(1+m)\tilde{f}} \quad (29)$$

$$\tilde{t}_0 = \frac{3p}{\tilde{f}^2}\left[\frac{1}{(2-1/n)\tilde{P}^{2(2-1/n)}} + \ln(\tilde{P}) + \tilde{f}\left(1-\frac{1}{\tilde{P}\tilde{f}}\right)\right] - \frac{6p}{(1+m)\tilde{f}} \quad (30)$$

Eqns. (29), (30) give the dependence of the dimensionless time $\tilde{t}$ as a function of $\tilde{X}$.

Let $t_1$ be the dimensionless time (in units of $t_0$) after which the tension has extended to all $N_{tot}$ monomers of the polymer. Then from Eqn.(24), we have

$$\tilde{X}(t_1) = N_{tot}{}^m \quad (31)$$

Substituting this equation into Eqn. (29) will yield an expression for $t_1$.

$$t_1 = \tilde{t}_0 - \frac{3p}{\tilde{f}^2}\left[\frac{1}{(2-1/n)\tilde{P}^{2(2-1/n)}} + \ln(\tilde{P}) + \tilde{f}\left(1-\frac{1}{\tilde{P}\tilde{f}}\right)\right]N_{tot}{}^{2m} + \frac{6pN_{tot}{}^{1+m}}{(1+m)\tilde{f}} \quad (32)$$

For time $\tilde{t} > t_1$, all $N_{tot}$ monomers of the polymer are now under tension. Hence Eqn. (25) now takes the form

$$\frac{\tilde{X}}{(2-1/n)\tilde{f}\tilde{P}^{2(2-1/n)}} + \frac{\tilde{X}}{\tilde{f}}\ln(\tilde{P}) + \tilde{X}(t)\left(1-\frac{1}{\tilde{P}\tilde{f}}\right) + M(\tilde{t}) = N_{tot}, \quad \tilde{t} \geq t_1 \quad (33)$$

Differentiating this equation with respect to $\tilde{t}$ and using Eqn. (27), we obtain

$$\left[\frac{1}{(2-1/n)\tilde{f}\tilde{P}^{2(2-1/n)}} + \frac{1}{\tilde{f}}\ln(\tilde{P}) + \left(1-\frac{1}{\tilde{P}\tilde{f}}\right)\right]\frac{d\tilde{X}}{d\tilde{t}} = -\frac{\tilde{f}}{6p\tilde{X}}, \quad \tilde{t} \geq t_1 \quad (34)$$

The solution of this equation is

$$\tilde{t} = t_1 + \frac{3p}{\tilde{f}^2}\left[\frac{1}{(2-1/n)\tilde{P}^{2(2-1/n)}} + \ln(\tilde{P}) + \tilde{f}\left(1 - \frac{1}{\tilde{P}\tilde{f}}\right)\right]\left[\tilde{X}(t_1)^2 - \tilde{X}(\tilde{t})^2\right] \quad (35)$$

where $\tilde{X}(t_1)$ is given by Eqn. (31). Eventually, at $\tilde{t} = t_{fin}$, the whole polymer exits from the nanopore, with $\tilde{X}(t_{fin}) = 0$. This gives

$$\tilde{t}_{fin} = t_1 + \frac{3p}{\tilde{f}^2}\left[\frac{1}{(2-1/n)\tilde{P}^{2(2-1/n)}} + \ln(\tilde{P}) + \tilde{f}\left(1 - \frac{1}{\tilde{P}\tilde{f}}\right)\right]\tilde{X}(t_1)^2 \quad (36)$$

If we denote by $t_2$ the dimensionless time (in units of $t_0$) as the time it takes, starting from $t_1$, for the polymer to exit the nanopore, then

$$t_2 = \frac{3p}{\tilde{f}^2}\left[\frac{1}{(2-1/n)\tilde{P}^{2(2-1/n)}} + \ln(\tilde{P}) + \tilde{f}\left(1 - \frac{1}{\tilde{P}\tilde{f}}\right)\right]N_{tot}^{2m} \quad (37)$$

The total translocation time $t = t_1 + t_2$. Using Eqns. (30), (32) and (37), the terms proportional to $N_{tot}^{2m}$ cancel and the translocation time is given by

$$t = \frac{3p}{\tilde{f}^2}\left[\frac{1}{(2-1/n)\tilde{P}^{2(2-1/n)}} + \ln(\tilde{P}) + \tilde{f}\left(1 - \frac{1}{\tilde{P}\tilde{f}}\right)\right] - \frac{6p}{(1+m)\tilde{f}} + \frac{6pN_{tot}^{1+m}}{(1+m)\tilde{f}} \quad (38)$$

which gives the translocation exponent as $1 + m$.

2. Weak force: $Ta/P^2 < f \leq T/P$

We will now consider the case when $f > Ta/P^2$, but less than $T/P$, where $P$ is the persistence length. Under the condition $f < T/P$, no blob of size less than or equal to $P$ can be formed. But a blob of size less than or equal to $P^2/a$ can still be formed. This is the so-called trumpet regime illustrated in Figure 1a.

The mass conservation equation is now replaced by

$$\int_{-X(t)}^{-X_Q(t)} \left[\frac{\mathbf{x}(x,t)}{a}\right]^{1/n} \frac{dx}{\mathbf{x}(x,t)} + \int_{-X_Q(t)}^{0} \left[\frac{\mathbf{x}(x,t)}{a}\right]^{2} \frac{dx}{\mathbf{x}(x,t)} + M(t) = N(t) \qquad (39)$$

The flux of monomers at the pore is $j_0(t) = \mathbf{r}(x=0,t)v(t)$, where $\mathbf{r}(x,t) = [\mathbf{x}(x,t)/a]^{1/n}/\mathbf{x}(x,t)$ is the linear density of monomers. But at the pore ($x=0$), where the force is the strongest, we have assumed that the monomers inside this blob are in the ideal chain state. Therefore $\mathbf{r}(0,t) = [\mathbf{x}(0,t)/a]^2/\mathbf{x}(0,t)$. The flux $j_0(t)$ should be equal to $dM(t)/dt$, i.e.

$$\frac{dM(t)}{dt} = \left[\frac{\mathbf{x}(x=0,t)}{a}\right]^2 \frac{v(t)}{\mathbf{x}(x=0,t)} = \frac{\mathbf{x}(x=0,t)v(t)}{a^2} \qquad (40)$$

Using Eqn.(6) with $z=3$, Eqn. (19), and introducing the dimensionless time $\tilde{t} = t/\mathbf{t}_0$, with $\mathbf{t}_0 = a^2 \mathbf{z}_0 / T$, this becomes

$$\frac{dM(\tilde{t})}{d\tilde{t}} = \frac{1}{6p\tilde{X}(\tilde{t})} \qquad (41)$$

Taking into account the above changes, with natural boundary condition $\tilde{X}(0) = 1/\tilde{f}$, and following the analysis for the case of intermediate force, we obtain for the translocation time

$$t = \frac{3p}{\tilde{f}^3}\left[\frac{1}{(2-1/n)\tilde{P}^{2(2-1/n)}} + \ln(\tilde{P}^2\tilde{f})\right] - \frac{6p}{(1+m)\tilde{f}^{1/m+1}} + \frac{6p}{(1+m)}N_{tot}^{1+m} \qquad (42)$$

The translocation exponent $1+m$ is the same as in the intermediate force case.

### 3. Strong force: $f > T/a$

In the strong force case, the entire part of the polymer under tension is completely stretched. This is illustrated in Figure 1c. The global balance of material now requires

$$\tilde{X}(t) + M(t) = N(t) \qquad (43)$$

At time $t=0$, the external force is zero and all $N(t)$ monomers have occupied the region of size $X(t)$. Therefore $N(t)$ must satisfy the condition (24), $N(t) = c\tilde{X}^{1/m}$ where $c$ is a constant and $m$ is an effective exponent for the end-to-end distance of the polymer. Therefore Eqn. (43) becomes

$$\tilde{X}(t) + M(t) = c\tilde{X}^{1/m} \qquad (44)$$

Differentiating this with respect to $\tilde{t}$ and using Eqn. (27) for $dM/d\tilde{t}$ yields

$$\left(1 - \frac{c}{m}\tilde{X}^{1/m-1}\right)\frac{d\tilde{X}}{d\tilde{t}} = -\frac{\tilde{f}}{6p\tilde{X}} \qquad (45)$$

The solution of this with initial condition $\tilde{X}(0) = 1$ is

$$\tilde{t} = \frac{6p}{\tilde{f}}\left(\frac{1}{2} - \frac{c}{1+m}\right) - \frac{3p}{\tilde{f}}\tilde{X}^2 + \frac{6p}{\tilde{f}}\frac{c}{1+m}\tilde{X}^{1/m+1} \qquad (46)$$

With these changes and following the analysis of the intermediate force case, we obtain the translocation time

$$t = \frac{6p}{\tilde{f}}\left(\frac{1}{2} - \frac{c}{1+m}\right) + \frac{6p}{\tilde{f}}\frac{c}{1+m}N_{tot}^{1+m} \qquad (47)$$

The constant $c < 1$ is necessary; otherwise, the first term can become negative. The dependence of the translocation time on the persistence length is only through the dependence of the effective exponent $m$ on $P$.

We have plotted the translocation time $t$ calculated using Eqns. (42) and (38) for weak force and intermediate force, respectively, in Figure 2 and Figure 3 respectively, as function of the total number of monomers $N_{tot}$. The effective exponent $m$ for the end-to-end distance of the polymer, as a function of $N_{tot}$ is approximated using the following function:

$$m = 0.5 + 0.1 \tanh^2((N_{tot} / \tilde{P}^2)^2) \qquad (48)$$

## III. Conclusion

We have studied theoretically the forced translocation of a semi-flexible polymer through a nanopore. Our method is a dynamic scaling approach based on the Pincus blob model in which the front of the tensile force propagates through the backbone of the polymer, as suggested by Sakaue [17]. This method was recently applied by Dubbledam et al [16, 18] to study the forced translocation of a completely flexible polymer with self-avoidance through a nanopore. We have generalized their method to study the case of a semi-flexible polymer with a persistence length $P$. We argue that semi-flexible polymer in translocation should include a self-avoiding region, a random walk region and a rigid rod region, under uneven tension propagation instead of uniform scaling picture. We propose that for a semi-flexible polymer under a sufficiently strong external force, two extra propagation fronts $X_Q$ and $X_P$ can be defined where the tensile blob size is equal to $P^2/a$ and the persistence length $P$, respectively, besides the force propagation front $X$ beyond which the force vanishes, such that $X > X_Q > X_P$. From $X$ to $X_Q$, the blob size $X$ size is larger than $P^2/a$ and the monomers inside such a blob obey self-avoidance statistics. From $X_Q$ to $X_P$, the blob size is smaller than $P^2/a$ and the monomers inside such a blob obey ideal chain statistics. From $X_P$ to 0, the blob size is smaller then $P$ and the monomers inside such a blob become a rigid rod. We study the model for the cases of weak, intermediate and strong external force $f$ and obtain analytic expressions for the translocation time $t$ for the three cases.

For weak external force $Ta/P^2 < f \leq T/P$, no blob of size less than $P$ can be formed but a blob of size $P^2/a$ can still be formed. The polymer under tension consists of a self-avoiding portion and an ideal chain portion. The translocation time as a function of the total number of monomers has the form $t = A(P,f) + \dfrac{6p}{(1+m)} N_{tot}^{1+m}$. For $N_{tot} \gg P^2/a$, the second term dominates. In that limit the translocation time becomes independent of $P$ and $f$ and the effective exponent $m = 3/5$. The translocation exponent is $a = 1+m = 1.6$. This is the same behavior as in a completely flexible chain. For $N_{tot} < P^2/a$, the translocation time depends on $P$ and $f$. The effective exponent $m = 0.5$ and the translocation exponent now has the value $a = 1.5$.

For intermediate external force $T/P < f \leq T/a$, a blob of size less than $P$ can also be formed. The polymer under tension consists of a self-avoiding portion, an ideal chain portion and also a rigid rod portion. The translocation time as a function of the total number of monomers has the form $t = B(P,f) + \dfrac{6pN_{tot}^{1+m}}{(1+m)\tilde{f}}$. For $N_{tot} \gg P^2/a$, the second term dominates. In that case the translocation time is independent of $P$ but decreases with force as $1/f$ and the translocation exponent is $a = 1.6$. This also agrees with the result for the completely flexible polymer. For $N_{tot} < P^2/a$, the translocation time depends on $P$ and $f$. The effective exponent $m = 0.5$ and the translocation exponent now has the value $a = 1.5$.

For the strong force case $f > T/a$, the part of the polymer under tension is completely stretched. The translocation time is now independent of $P$ and has the form

$$t = \frac{C}{\tilde{f}} + \frac{6pN_{tot}^{1+m}}{(1+m)\tilde{f}}.$$

For $N_{tot} \gg P^2/a$, the second term dominates and it has the same form as in the case of intermediate force, with translocation $a = 1.6$. For $N_{tot} < P^2/a$, the translocation time is independent of $P$ and decreases with force as $1/f$, but with translocation exponent $a = 1.5$.

In spite of the complexity of the technical analysis, the final results of the translocation time turn out to be rather simple. For the case of weak force, $t \propto N_{tot}^{1+m}$, independent of external force. For intermediate and strong forces, it is inversely proportional to the external force: $t \propto \frac{N_{tot}^{1+m}}{f}$. In all cases the effective exponent $m$ varies between ½ and 3/5, depending on the total length of the polymer. This effective exponent enters mainly through Eqn. (24) which relates the total number of monomers subject to tension $N(t)$ to $X(t)$. Eqn. (24) holds because at $t = 0$, the external force is zero and all $N(t)$ monomers must have occupied the region of size $X(t)$.

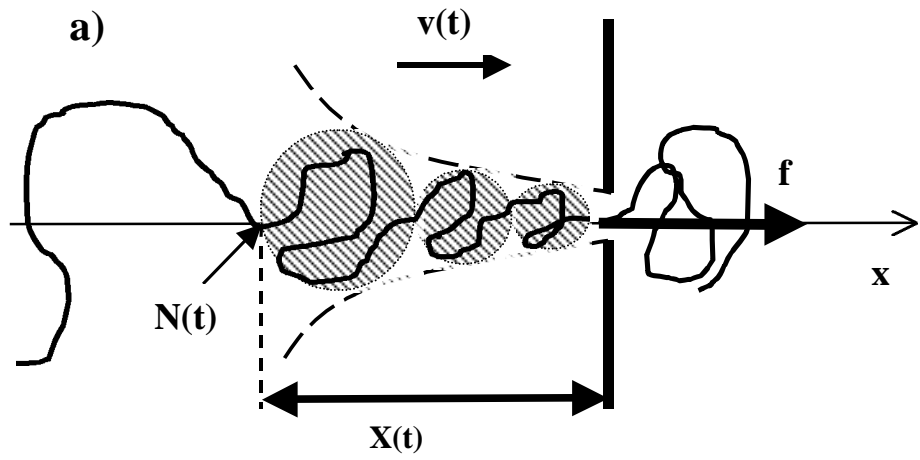

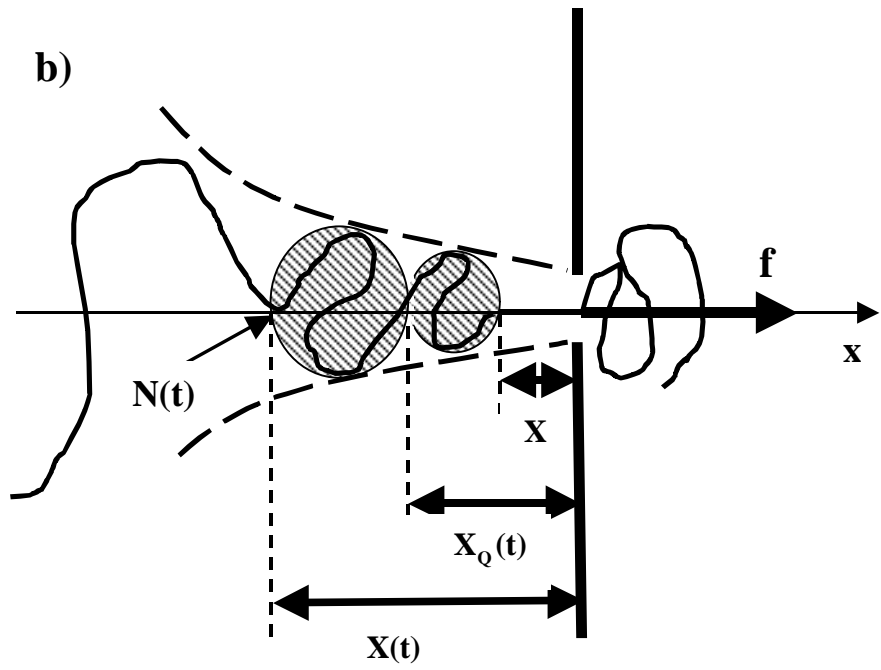

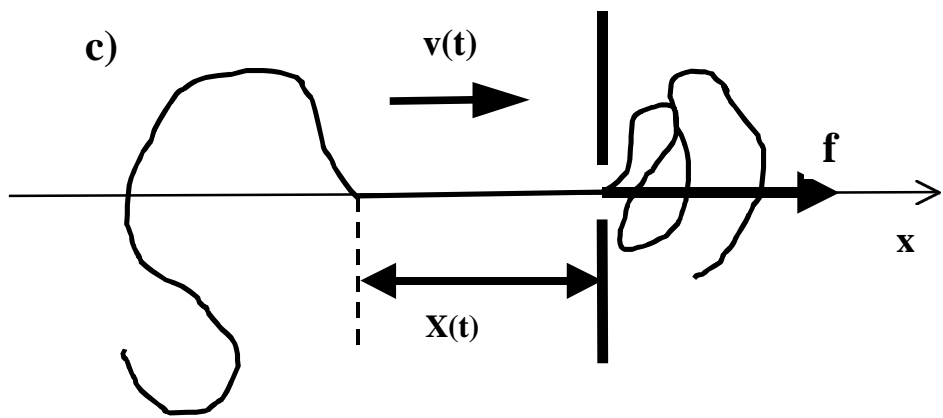

Figure 1. Configurations of the polymer under a driving force acting at the pore. For weak force $Ta/P^2 < f \leq T/P$, the polymer on the left side of the pore form tensile blobs in the shape of a "trumpet" (a). For intermediate force $T/P < f < T/a$, the part of the polymer near the pore up to $x = -X_P$ is completely stretched by the force, while further away, tensile blobs are still formed. This is the "stem-trumpet" regime (b). In the strong force regime, $f > T/a$, all of the polymer under tension $X(t)$ is completely stretched (c).

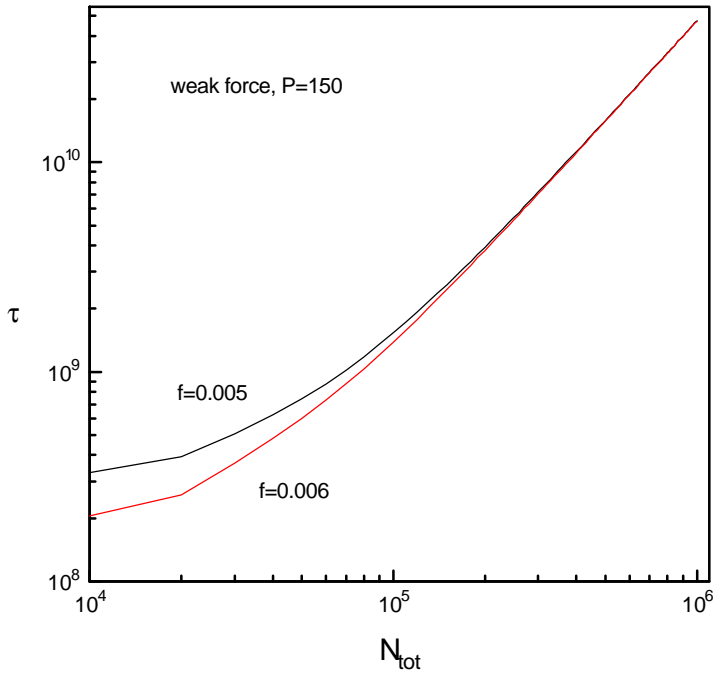

Figure 2. Translocation time $t$ plotted versus the total number of monomers for various values of the dimensionless external force $\tilde{f}$, for the case of weak force. The persistence length is fixed at $\tilde{P} = 150$.

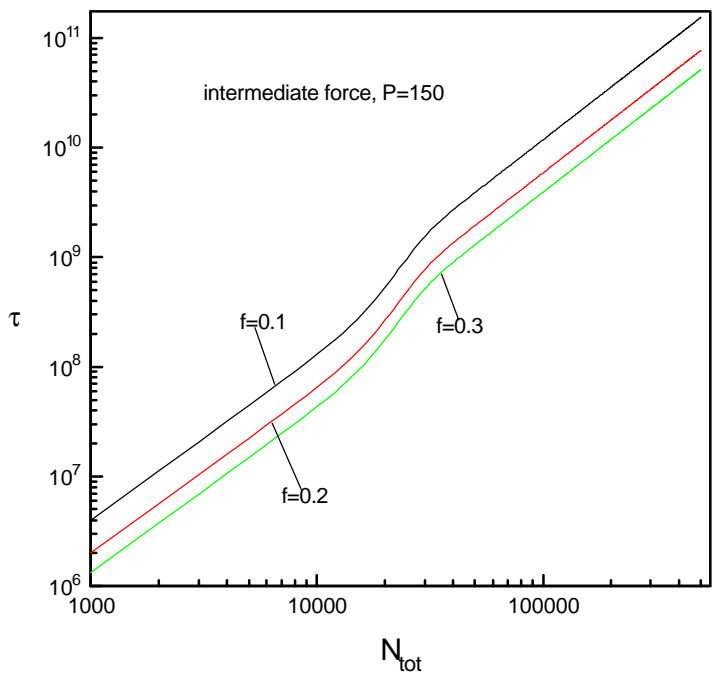

Figure 3. Translocation time $t$ plotted versus the total number of monomers for various values of the dimensionless external force $\tilde{f}$, for the case of intermediate force. The persistence length is fixed at $\tilde{P} = 150$.